\documentclass[letterpaper]{jpconf}
\usepackage{graphicx}

\begin{document}

\hfill MAN/HEP/2010/6

\title{A measurement of the muon charge asymmetry in W boson events}

\def\etal{{\sl et al.}}
\newcommand{\met}{\mbox{$E\kern -0.6em/_{\rm T}$}}
\newcommand{\ptmu}{\mbox{$p_{T}^{\mu}$}}

\author{Mika Vesterinen on behalf of the D0 Collboration}

\address{Particle Physics Group, School of Physics and Astronomy, University of Manchester, UK.}

\begin{abstract}
A recent measurement by the D0 Collaboration, of the charge asymmetry of muons in 
$W\rightarrow\mu\nu$ events is discussed.
Using 4.9 fb$^{-1}$ of data collected by the D0 detector, the muon charge asymmetry, $A(\eta)$, 
where $\eta$ is the muon pseudorapidity, is measured inclusively and in two bins of 
muon transverse momentum, \ptmu.
After being corrected for background contamination and detector effects (resolution and acceptance),
the data are compared to NLO QCD predictions with {\sc CTEQ} 6.6 PDFs.
For inclusive \ptmu, reasonable agreement is observed.
However, the prediction fails to describe the data in bins of \ptmu.
In most $\eta$ bins, the data are more precise than the predictions,
and will provide tight constraints on future PDFs.

\end{abstract}

\section{Introduction}

Studies of $W$ and $Z$ boson production at hadron colliders,
provide valuable insight into the QCD production mechanism.
The decay channels, $W\rightarrow l\nu$ and $Z\rightarrow ll$ ($l = e,\mu$),
offer colourless final states, whose kinematic distributions probe 
the parton distribution functions (PDFs) of the colliding hadrons.
More precise PDFs will ultimately improve our sensitivity to new physics
signals at hadron colliders.

In $p\bar{p}$ collisions, 
$W$ bosons are mostly produced via the scattering of valence $u$ or $d$ quarks from the proton,
with the corresponding isospin partner antiquark from the antiproton.
Since $u$($\bar{u}$) quarks carry more of the proton(antiproton) momentum than $d$($\bar{d}$) quarks,
$W^+$'s tend to be boosted along the proton direction, and $W^-$'s along the antiproton direction.
This production asymmetry, $A(y)$, is defined as

\begin{equation}\label{Ay} 
A(y) = \frac{d\sigma(W^+)/dy - d\sigma(W^-)/dy}{d\sigma(W^+)/dy + d\sigma(W^-)/dy} 
\end{equation}
where $y$ is the $W$ boson rapidity.
Unfortunately, $y$, and thus also $A(y)$ are not directly measurable in $W\rightarrow l\nu$ decays, 
due to the unknown momentum of the neutrino along the beam direction.
However, given an additional constraint to the known $W$ boson mass,
it is possible to estimate $y$ on a statistical basis,
assuming the well understood $V-A$ decay coupling~\cite{CDF_method}.
This analysis technique was adopted by the CDF collaboration in a recent
measurement of $A(y)$,
with 1 fb$^{-1}$ of data in the electron channel~\cite{CDF_AW}.

A directly measurable observable, which is also sensitive to $A(y)$, 
is the analogous lepton charge asymmetry, $A(\eta)$ defined as
\begin{equation}\label{Aeta} 
 A(\eta) = \frac{d\sigma(\mu^+)/d\eta - d\sigma(\mu^-)/d\eta}{d\sigma(\mu^+)/d\eta + d\sigma(\mu^-)/d\eta}
 \end{equation}
where $\eta$ is the lepton pseudorapidity.
$A(\eta)$ is less directly sensitive to the PDFs, since it is a convolution of the production asymmetry with 
the decay asymmetry of the $V-A$ coupling in the decay vertex;
the two of which tend to cancel.
However, the production asymmetry dominates at small $\eta$ and/or large \ptmu,
hence the sensitivity to PDFs is enhanced by further binning in \ptmu.
The D0 collaboration has measured $A(\eta)$ with 300 pb$^{-1}$ in the muon channel~\cite{D0_AMU}, and 700 pb$^{-1}$ in the
electron channel~\cite{D0_AE}.

This document describes a measurement of $A(\eta)$ in the muon channel, using 4.9 fb$^{-1}$ of data~\cite{D0_AMU_CONF},
which supersedes~\cite{D0_AMU}.

\section{The D0 detector}

The D0 detector is detailed in~\cite{D0_NIM}.
Here, only the features that are essential to this measurement 
are described.
Closest to the interaction region,
an inner tracker covering the region $|\eta|$ $<$ 3,
and within a 2 T solenoidal magnet
determines the momentum and charge of muons.
A liquid argon and uranium calorimeter is contained within three separate cryostats; 
it determines the missing transverse energy of the event,
and provides isolation information for muon identification purposes.
Finally, a muon spectrometer covering the region $|\eta|$ $<$ 2, with wire chambers, and trigger scintillator counters,
either side of a 1.8 T toroidal magnet identifies muon candidates.
An important feature of the D0 detector,
is that both magnet polarities are reversed periodically, 
such that the analysed data set contains roughly equal luminosities
for each of the four combinations. 
This dramatically reduces the effect of any charge asymmetry in muon
reconstruction efficiencies.

\section{Event selection}

Candidate $W\rightarrow\mu\nu$ events are required to fire a single muon trigger
and contain one muon reconstructed in the muon spectrometer, which is also spatially matched to a
track in the inner tracker with $p_T$ $>$ 20 GeV.
Timing cuts based on muon scintillator information reject cosmic ray muons.
Additionally, the muon candidate must be isolated in the calorimeter and inner tracker in order to reject
jets that fake muons, or semi-leptonic decays of hadrons within jets.
The missing transverse energy, \met, must be greater than 20 GeV, consistent with a high transverse momentum
neutrino, and the transverse mass, 
$M_T$ = $\sqrt{2p_T^{\mu}\met(1-\cos\phi)}$, where $\phi$ is the 
azimuthal opening angle between the muon and the neutrino,
must be greater than 40 GeV.

\section{Detector corrections and background subtraction}

The measured $A(\eta)$ must be corrected for the following:
\begin{enumerate}
\item Detector resolution and acceptance,
\item Backgrounds,
\item Charge mis-identification.
\end{enumerate}

In order to determine corrections for detector resolution and acceptance, 
$W\rightarrow\mu\nu$ signal events are generated using {\sc pythia}~\cite{pythia}, and passed through a detailed {\sc geant}~\cite{geant} 
based simulation of the D0 detector,
before being overlaid with randomly triggered events from data.
The simulation is further corrected for differences compared to data, 
in muon reconstruction efficiencies and momentum resolution.
Systematic uncertainties on the detector corrections are estimated by varying the 
momentum resolution in the simulation within its uncertainty.

Backgrounds from $W\rightarrow\tau\nu$, $Z\rightarrow\mu\mu$ and $Z\rightarrow\tau\tau$ processes,
are simulated in the same way as the signal, and are subtracted from the data.
Multijet backgrounds are estimated directly from data.
The efficiency for a jet to fake a muon, $f$,  is estimated from a sample of muon candidates which are
back-to-back in azimuth with a reconstructed jet, that also fired a jet trigger.
The missing transverse energy of the event is required to be less than 10 GeV, 
in order to suppress $W\rightarrow\mu\nu$ signal.
Systematic uncertainties on $f$, are estimated by adding requirements that further suppress
signal contamination.

The charge mis-identification rate is estimated using $Z\rightarrow\mu\mu$ candidate events, 
and is found to have negligible effect on the measurement.

\section{Results}

\begin{figure}
\centering
\includegraphics[width=0.60\textwidth]{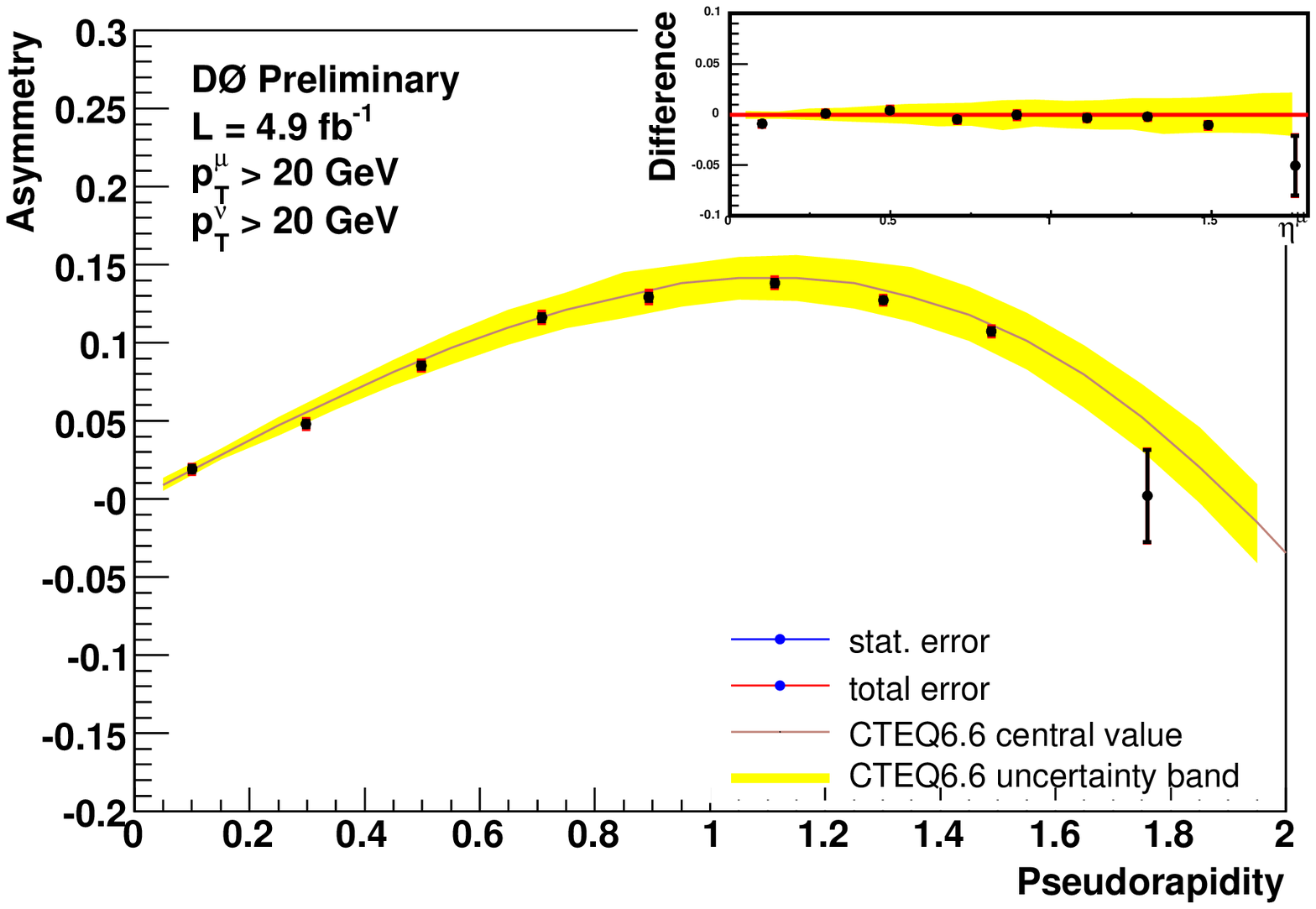}\\
\includegraphics[width=0.60\textwidth]{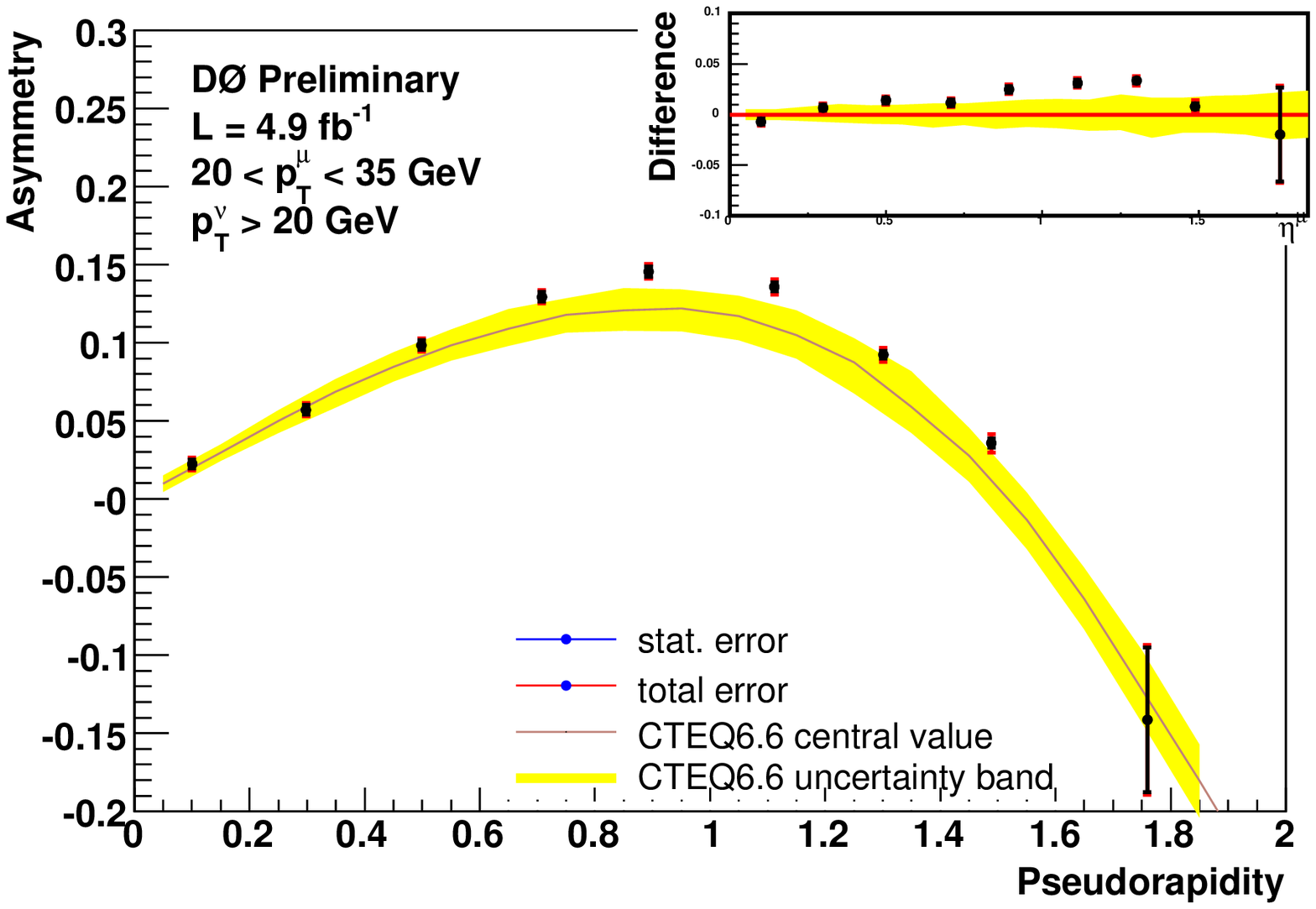}
\includegraphics[width=0.60\textwidth]{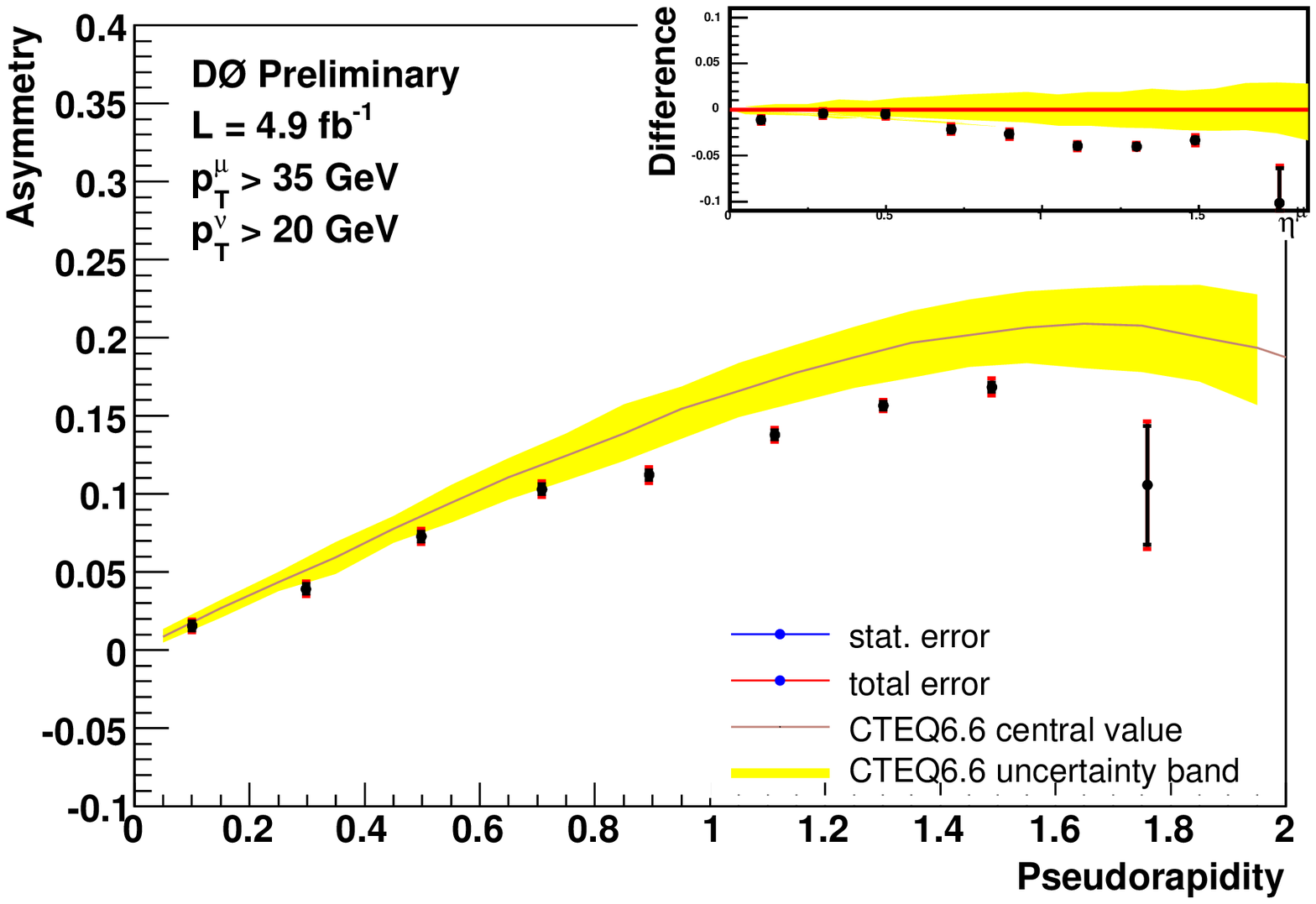}
\caption{Comparison of the measured $A(\eta)$ with NLO QCD predictions for
(top) \ptmu\ $>$ 20 GeV, (middle) $20 < \ptmu < 35$ GeV, and (bottom) \ptmu\ $>$ 35 GeV.}
\label{Figure:results}
\end{figure}

After CP folding (assuming $A(\eta) = -A(-\eta)$), figure~\ref{Figure:results} compares the measured $A(\eta)$ 
with NLO QCD predictions from 
from {\sc resbos}~\cite{resbos} using {\sc cteq} 6.6 PDFs~\cite{cteq66},
with QED radiative corrections from {\sc photos}~\cite{photos}.
The yellow bands represent the PDF uncertainties on the predictions.
For inclusive \ptmu, the data and prediction agree well.
However, the agreement is rather poor once the measurement is split into bins
of \ptmu.

\section{Conclusions}

The charge asymmetry of muons from from $W\rightarrow\mu\nu$ decays, $A(\eta)$,
has been measured using 4.9 fb$^{-1}$ of data collected by the D0 detector.
Predictions from NLO QCD with {\sc cteq} 6.6 PDFs are able to describe 
the data for inclusive muon \ptmu, but rather poor agreement with the data is observed
in bins of \ptmu.
Apart from the highest $\eta$ bins, the data are more precise than the predictions,
and will thus provide tight constraints on future PDFs.

\end{document}